# Emergence of mutationally robust proteins in a microscopic model of evolution

Konstantin B. Zeldovich and Eugene I. Shakhnovich

Department of Chemistry and Chemical Biology, Harvard University,
12 Oxford St, Cambridge, MA
E-mail: eugene@belok.harvard.edu

The ability to absorb mutations while retaining structure and function, or mutational robustness, is a remarkable property of natural proteins. In this Letter, we use a computational model of organismic evolution [Zeldovich et al, *PLOS Comp Biol* 3(7):e139 (2007)], which explicitly couples protein physics and population dynamics, to study mutational robustness of evolved model proteins. We find that dominant protein structures which evolved in the simulations are highly designable ones, in accord with some of the earlier observations. Next, we compare evolved sequences with the ones designed to fold into the same dominant structures and having the same thermodynamic stability, and find that evolved sequences are more robust against point mutations, being less likely to be destabilized upon them. These results point to sequence evolution as an important method of protein engineering if mutational robustness of the artificially developed proteins is desired. On the biological side, mutational robustness of proteins appears to be a natural consequence of the mutation-selection evolutionary process.

Present-day proteins are the result of billions of years of evolution, whereby mutations in DNA induce a fitness changes in its carrier organism, and are eventually fixed in or removed from the population. Mutational robustness is one of the most striking properties of evolved natural proteins. In addition to maintaining viability of organisms in the presence of mutations, mutational robustness increases the number of sequences surviving within a given scaffold structure, and thus promotes sequence diversification and "evolvability" [1].

Not surprisingly, substantial analytic and simulation effort has been devoted to the study of mutational robustness and its origins and consequences. For example, Taverna and Goldstein [2] considered a lattice protein model and evolved sequences according to mutation and selection scheme, with viable sequences defined as having the stability or folding free energy $\Delta G$ lower than a certain cutoff $\Delta G_{crit}$. They have shown that more stringent stability requirement $\Delta G_{crit}$ made evolved sequences more robust against point mutations. In contrast, random sequences were never robust. However, this insightful paper did not consider the possible effect of structure designability on the mutational robustness of both evolved and random sequences. In particular, these authors did not rule out a possibility that increased mutational robustness of evolved proteins is a straightforward consequence of their folding into highly designable structures, attracting larger volumes of sequence space.

Recently, we developed a computational model of early evolution [3] which takes into account most key biological processes on both molecular (mutations and gene duplication) and organismal (replication and death) levels and makes an explicit, justifiable connection between the two via the dependence of fitness of an organism on the stability of its proteins. Assuming that all proteins must be in their native states in a viable organism, we suggested that the organism death rate increases as the stability of its proteins decreases. This simple assumption, coupled with an explicit calculation of protein stability in a physics-based lattice model and simplest population dynamics, allowed us to make quantitatively correct predictions of the sizes of evolved protein families and superfamilies, to reproduce the scale-free nature of the protein domain universe graph [4], and to hypothesize that rapid exponential growth of the protocell population is intricately linked with the discovery of a few dominant protein structures (DPS) that attract the majority of evolving sequences and dominate the structural repertoire of the evolving genes. Among the key questions that remained unanswered in the original publication, however, was the mysterious nature of the DPS discovered in the simulations: DPS did not exhibit any apparent structural difference from non-DPS or similarity between themselves. The issue of mutational robustness was not considered at all. One of the reasons for these limitations was the computational effort required to calculate the properties of evolved proteins based on a complete library of 103346 compact 27-mer structures [5]. In this paper, we use a reduced representative set of 10000 structures for greater computational efficiency and find that DPS structures are in fact the highly designable ones. Further, and most importantly, by comparing the response of evolved and designed sequences to point mutations, we show that our evolutionary model produces sequences that are more evolvable and have a higher mutational robustness than sequences with identical stability and native structure, but obtained by traditional sequence design methods.

In the model of organismic evolution [3], an organism is represented by a set of several (typically, below 10) 81-nucleotide genes, encoding 27-mer proteins. Organisms can replicate and die, and genes within each organism can mutate or duplicate. The mutation rate (a random change of a randomly chosen nucleotide, with mutations to stop codons explicitly rejected) is $m=0.3$ per gene per timestep, and the gene duplication rate is $g=0.03$. All organisms start with one random gene, and undergo the four elementary events (replication, death, mutation, gene duplication) described above. After each mutation, a 3x3x3 lattice model [6] is used to determine the native structure and thermodynamic stability of each protein. To do so, we calculate the energy $E_i$ of the sequence in each of the 10000 permitted conformations according to a contact pairwise potential derived by Miyazawa and Jernigan [7]. The set of 10000 structures has been selected randomly from the complete set of 103346 lattice 27-mers, and remained fixed for all simulations. Protein stability $P_{nat}$, or the thermodynamic probability for the protein to be in its native state at temperature $T=0.8$ is then calculated according to

$P_{nat} = \frac{\exp(-E_0/T)}{\sum_{i=0}^{10000} \exp(-E_i/T)}$, where $E_0$ is the energy of the sequence in its native (lowest-energy) conformation. Organism fitness is linked to its proteins' stability according to the "weakest link" hypothesis [3], whereby the death rate of the organism is defined by the stability of the least stable protein in its genome: $d = d_0(1 - \min_{i=1..N} P_{nat}^{(i)})$. This relationship ensures that organisms with less stable proteins have a higher death rate, thus lower fitness, and are gradually removed from the evolving population. The birth, or replication rate of the organisms is constant, $b=0.15$, and the initial death rate $d_0$ is chosen in such a way that for $d=b=0.15$ for the initial organisms with a single random gene. The population of model organisms evolves for up to 10000 time steps; whenever the population size exceeds 5000, excess organisms are randomly removed from the population, similar to a chemostat experiment.

We estimate the designability of a structure as the number of sequences (maybe having a certain property, such as stability) folding into a given structure [8] [9, 10]. To estimate the designability of each of the 10000 lattice structures, we created a set of $10^7$ random 27-mer sequences, determined their native states, and defined the designability of each structure as the number of sequences having the structure as the native state. Then, we sorted the list of structures according to their designability, and introduced the designability rank $D^{rand}$ as the position of the structure in the list, with the lowest rank of $D^{rand}=1$ corresponding to the least designable structure, and the highest rank $D^{rand}=10000$ corresponding to the most designable one. The average stability $P_{nat}$ of the random sequences at $T=0.8$ was 0.268. We repeated the same procedure for a set of $10^7$ sequences that underwent Monte Carlo sequence optimization [11] to achieve average stability of $P_{nat}=0.707$, and determined the designability rank $D^{des}$ for this set of designed sequences. The designability ranks $D^{rand}$ and $D^{des}$ of a structure are highly correlated ($R=0.98$), showing that designability rank of a structure is essentially independent of the average stability of the sequences used to estimate it.

We ran 200 independent evolution simulations over 10000 timesteps (~1400 generations) each, resulting in 163 survivals and 37 extinctions of the population. The typical mutation rate in surviving organisms was about 1 mutation per gene per generation. In the surviving populations, we defined the dominant protein structure (DPS) as a structure serving as the native state to the largest number of sequences in each populations. In total, we identified 163 DPS corresponding to about $1.75 \cdot 10^6$ sequences. Some of the DPS have been discovered convergently, in unrelated runs, resulting in 120 distinct structures. The dominant structures encompassed about 69% of all sequences present in all surviving organisms, pointing out at their especially favorable structural properties. As it is not obvious which particular properties make a protein structure evolutionarily favorable, we decided to investigate designability rank of the dominant structures. The designability rank of a structure, from 1 to 10000, is an indicator of the number of sequences a structure can potentially attract. In Figure 1, we present the histogram of the designability ranks $D^{rand}$ for the 120 distinct DPS discovered in the simulations. As one can see, 119 of the DPS have the designability rank of 8000 or above, with the only having a lower rank of 4322. Thus, the absolute majority of the DPS are highly designable structures. This result is in quantitative agreement with an earlier finding of Taverna and Goldstein [12] who reported "enrichment" of highly designable structures by sequences found in a mutation-selection scheme.

One could argue, however, that the measure of designability may depend on the set of sequences used to estimate designabilities in a set of structures. For example, the ensemble of random sequences was used to estimate designability in [10], whereas real proteins obviously correspond to very non-random sequences possessing at least high thermodynamic stability. Therefore, *a priori* it is not evident that a structure serving as the native state for a large number of stable sequences will be as efficient for attracting random sequences. To demonstrate that designability of a lattice structure is robust with respect to the stability of protein sequences, we estimated designability of each structure based on two sets of sequences, random sequences with average stability $P_{nat}$=0.268, and stable, designed [11, 13] sequences with average stability $P_{nat}$=0.707. In Figure 2, we plot the designability rank $D^{des}$ of a structure calculated over the set of designed sequences, as function of its designability rank $D^{rand}$ for the random sequences (black dots). The two designability ranks are very strongly correlated, proving that designability of a structure is not sensitive to the average stability of sequences under consideration. The 120 DPS discovered in the evolution simulations are shown in Figure 2 as blue circles. They congregate in the upper right corner of the plot, further demonstrating that highly designable structures are the likely candidates for dominance in the course of protein evolution.

We note that our initial finding [3] that designability does not play a big role in the structural selection of potential DPS, is most likely due to an inadequate sampling of the space of 103,346 different structures by just 27 successful evolution runs.

With high designability established as the key structural aspect of DPS, we now turn to the investigation of sequences which evolved to fold into the DPS, in particular of their mutational robustness. As most of the sequences within each evolution run have resulted from divergent evolution and therefore are homologous, we applied the minimal Hamming distance cutoff of 14 within each run to distill the pool of $1.75 \cdot 10^6$ DPS-folding sequences to just 50077 nonhomologous ones. The average stability of these sequences was $<P_{nat}>$=0.808. We then attempted all possible aminoacid mutations in each of the 50077 DPS-folding, nohomologous sequences, and calculated the change of stability

$\Delta P_{nat}$ (a lattice analog of $\Delta\Delta G$, change in folding free energy upon point mutation in real proteins) upon each of the mutations. The distribution of $\Delta P_{nat}$ for evolved DPS-folding sequences is shown in Figure 3, red curve. To see whether this distribution is in any way singular, we used a Monte Carlo sequence design procedure [11] to create a set of 50000 designed sequences with the same average stability $<P_{nat}>=0.80$. Starting from a random sequence, the Monte Carlo procedure accepted mutations increasing stability $P_{nat}$ until the required average stability of the native state was reached. Note that the design procedure did not make any assumptions about the native states of the structures; after each mutation, the native state was determined exactly as the one having lowest energy among 10000 conformations allowed by the lattce model. These designed sequences were nonhomologous to each other, so Hamming distance filtering was not used. Then, designed sequences were subject to all possible mutations, and the distribution of $\Delta P_{nat}$ is shown in Figure 3, black curve. One can immediately see that on average, evolved sequences have a smaller average decrease in stability upon point mutation than designed ones (-0.104 vs. -0.144) *despite having the same stability*.

One can argue, however, that the difference in mutation responses could be attributed to the differences between the structures of the evolved DPS and the (highly designable, in fact) native states of designed sequences. To reject this possibility, we used Monte Carlo sequence design to create another set of 50077 sequences having exactly the same native states as the DPS, same number of sequences per each of the 120 distinct DPS, and same average stability as the DPS-folding evolved nonhomologous sequences. The green curve in Figure 3 shows the distribution of $\Delta P_{nat}$ after point mutations in these sequences, fully controlled for stability and structure. Being equal in the most obvious respects, stability and native states, the designed sequences are still much less robust to point mutations than evolved ones (average $\Delta P_{nat}$ of -0.138 vs. -0.104). Also, the probability to find a strongly stabilizing mutation is much higher in the evolved sequences – the right tail of the red plot in Figure 3 is way above the black and green ones. This finding shows higher evolvability of DPS-folding sequences.

To verify the robustness of the model, we have repeated all simulations using the complete set of 1081 25-mers on a two-dimensional 5x5 lattice as a protein model. All predictions (data not shown), including the key features of the mutation robustness plot, Figure 3, were identical to the results of the 3x3x3 27-mer model.

These results unequivocally point to the subtle but very important differences between designed and evolved sequences: whereas one can use sequence design to create sequences with desired stability and native structure, the designed sequences may be inherently unstable against point mutations. In contrast, sequence evolution, proceeding by multiple rounds of mutation, replication, and selection, allows one (or the Nature) to create entire families of stable, mutationally robust sequences.

Based on a simple, bottom-up model of molecular evolution, coupling protein physics with population dynamics, we have demonstrated the fundamental difference between protein evolution and protein design. In protein design, one is typically interested in finding a sequence with a desired native state and reasonably high stability against unfolding. The physics-based design procedures introduce mutations in the candidate sequences and optimize them until the native state satisfies both the thermodynamic and structural constraints. While the result may seem satisfactory, our finding suggests that it can be of lesser biological relevance: the designed sequence may be not at all robust against mutations, and introducing mutational robustness as an explicit design requirement remains a very challenging task. A natural solution to the problem is to pass from sequence optimization to sequence evolution algorithm, mimicking the natural processes of mutation, replication, and selection. In this case, stability and mutational robustness are inherently coupled. Indeed, even a very stable, but not mutationally robust sequence will not survive the multiple rounds of mutation and selection.

The most direct way of testing our predictions experimentally would be to introduce mutations in an artificially designed protein and compare their effect with that in a similar natural protein. Unfortunately, at present this approach does not appear to be feasible. For example, although entire families of designed proteins have been reported [14], these have been created by statistical methods mimicking the sequence alignment of the natural protein families, and not by physics-inspired procedure. On the contrary, Top7, a novel artificial protein [15] designed with extensive use of physical insight in the structure prediction, lacks well-studied natural structural analogs. Future experimental work will be required to compare the mutational robustness of biological and designed proteins.

Aside from the technical implications for protein design, our demonstration of the role of evolution in developing mutational robustness is of biological significance. Indeed, it shows that mutational robustness naturally develops in a very simple, physics-based model of evolution, where the only ingredients are the sequence-structure relationship of protein physics and the genotype-phenotype selection feedback loop of population dynamics. It is very likely that mutational robustness observed in present-day natural proteins is a distant echo of the simple, violent mutation-selection processes, akin to our simulations, which probably occurred at the early origins of Life.

**Acknowledgments**
This work was supported by the NIH. We are grateful to S. Wallin for stimulating discussions and to R. Olivares-Amaya for help with simulations.

## References

[1]   D. J. Earl, and M. W. Deem, Proc Natl Acad Sci U S A **101**, 11531 (2004).
[2]   D. M. Taverna, and R. A. Goldstein, J Mol Biol **315**, 479 (2002).
[3]   K. B. Zeldovich, Chen, P., Shakhnovich, B. E., Shakhnovich, E.I., PLoS Comp Biol **3**, e139 (2007).
[4]   N. V. Dokholyan, B. Shakhnovich, and E. I. Shakhnovich, Proc Natl Acad Sci U S A **99**, 14132 (2002).
[5]   E. I. Shakhnovich, and A. Gutin, J Chem Phys **93**, 5967 (1990).
[6]   E. Shakhnovich, Gutin, A, Journal of Chemical Physics **93**, 5967 (1990).
[7]   S. Miyazawa, and R. L. Jernigan, J Mol Biol **256**, 623 (1996).
[8]   A. V. Finkelstein, A. M. Gutin, and A. Badretdinov, Subcell Biochem **24**, 1 (1995).
[9]   H. Li *et al.*, Science **273**, 666 (1996).
[10]  H. Li, C. Tang, and N. S. Wingreen, Proteins **49**, 403 (2002).
[11]  I. N. Berezovsky, K. B. Zeldovich, and E. I. Shakhnovich, PLoS Comput Biol **3**, e52 (2007).
[12]  D. M. Taverna, and R. A. Goldstein, Biopolymers **53**, 1 (2000).
[13]  M. P. Morrissey, and E. I. Shakhnovich, Fold Des **1**, 391 (1996).
[14]  W. P. Russ *et al.*, Nature **437**, 579 (2005).
[15]  B. Kuhlman *et al.*, Science **302**, 1364 (2003).

**Figure captions.**

**Figure 1.**
Histogram of designability ranks $D^{rand}$ of the evolved dominant protein structures (DPS). 119 of the 120 evolved DPS are in the top 20% of most designable structures.

**Figure 2.**
Designability rank of structures $D^{des}$ calculated over designed sequences with stability $P_{nat}$=0.707 is highly correlated with the designability rank $D^{rand}$ of the same structures, calculated over unstable, random sequences. The DPS (blue circles) are in the top right corner of the diagram, and are highly designable according to either definition of designability.

**Figure 3.**
Mutational robustness of evolved and designed sequences: Distribution of changes of stability $\Delta P_{nat}$ upon point mutations for nonhomologous evolved DPS-folding sequences (red, average stability $P_{nat}$=0.808), designed sequences with the same stability (black), and designed sequences with same stability and native states as the evolved ones (green). The average change of stability upon point mutation is $<\Delta P_{nat}>$=-0.104 for evolved sequences, $<\Delta P_{nat}>$=-0.144 for designed sequences with same stability, and $<\Delta P_{nat}>$=-0.138 for designed sequences with same stability and native states. Negative values of $<\Delta P_{nat}>$ mean that, on average, proteins are destabilized by point mutations. Having the same stability (or even the same native state) as the designed sequences, evolved model proteins are more mutationally robust (smaller magnitude of $<\Delta P_{nat}>$ ) and are more evolvable, i.e. more susceptible to stabilizing mutations.

**Figure 1**

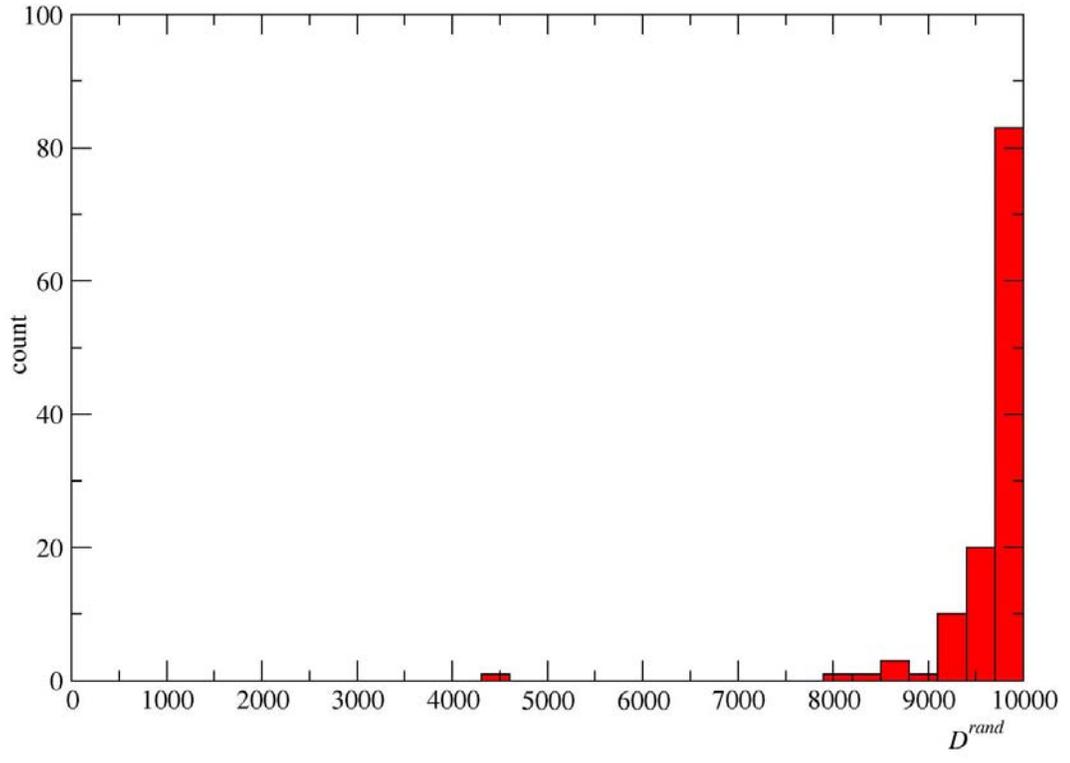

Figure 2

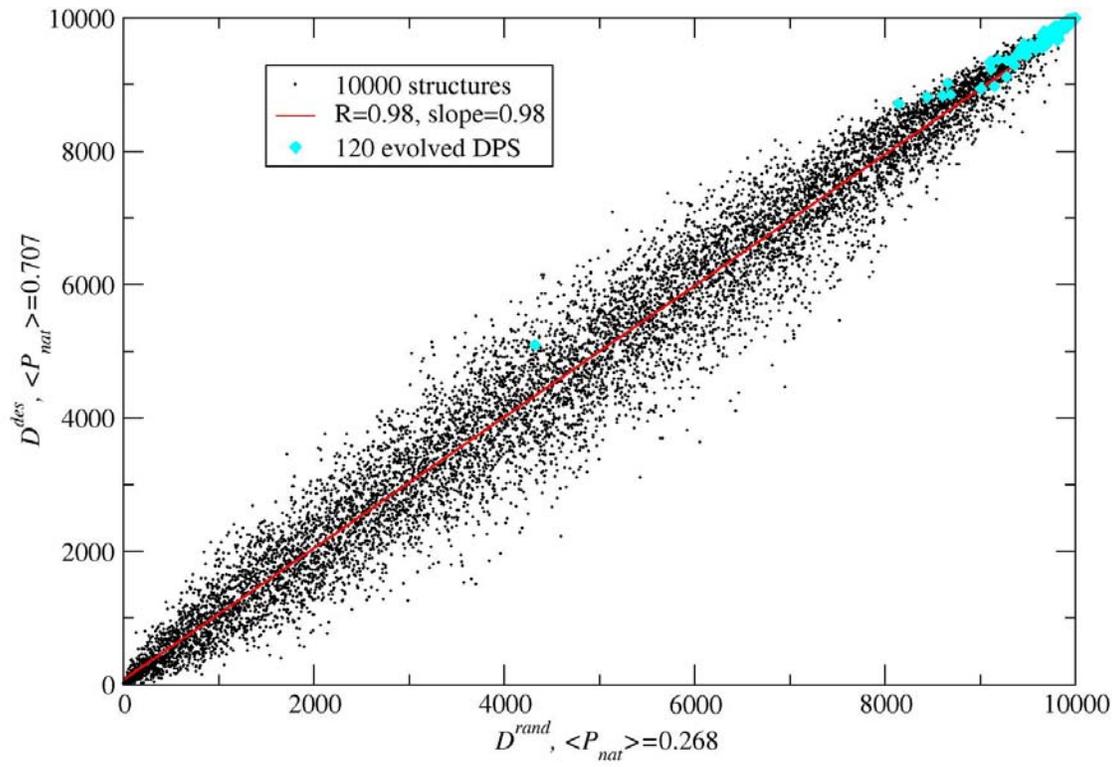

Figure 3

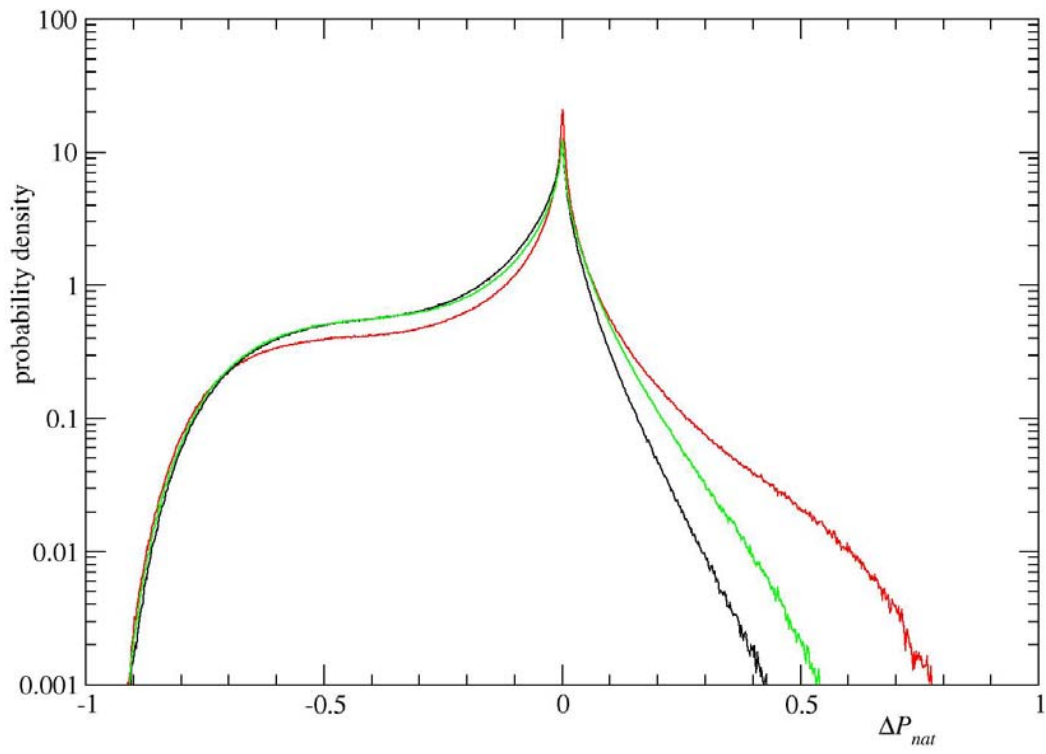